\documentclass[10pt]{wlscirep}
\usepackage[utf8]{inputenc}
\usepackage[T1]{fontenc}

\usepackage{graphicx}
\usepackage{amsmath}
\usepackage{amssymb}
\usepackage{bm}
\usepackage{adjustbox}
\usepackage{cleveref}
\crefname{equation}{Eq.}{Eqs.}  
\Crefname{equation}{Equation}{Equations}	
\crefname{figure}{Fig.}{Figs.}
\Crefname{figure}{Figure}{Figures}
\crefname{chapter}{Ch.}{Chs.}
\Crefname{chapter}{Chapter}{Chapters}
\crefname{section}{Sec.}{Secs.}
\Crefname{section}{Section}{Sections}
\crefname{appendix}{App.}{App.}
\Crefname{appendix}{Appendix}{Appendices}	
\crefname{algorithm}{Alg.}{Algs.}
\Crefname{algorithm}{Algorithm}{Algorithm}
\crefname{table}{Table}{Tables}
\Crefname{table}{Table}{Tables}
\let\originalleft\left
\let\originalright\right
\renewcommand{\left}{\mathopen{}\mathclose\bgroup\originalleft}
\renewcommand{\right}{\aftergroup\egroup\originalright}

\title{Interlayer excitons in van der Waals heterostructures: Binding energy, Stark shift, and field-induced dissociation}

\author[1,2,*]{H{\o}gni C. Kamban}
\author[1,2]{Thomas G. Pedersen}
\affil{Department of Materials and Production, Aalborg University, DK-9220 Aalborg \O st, Denmark }
\affil[2]{Center for Nanostructured Graphene (CNG), DK-9220 Aalborg \O st, Denmark}

\affil[*]{hck@mp.aau.dk}

\begin{abstract}
Photoexcited intralayer excitons in van der Waals heterostructures (vdWHs) with type-II band alignment have been observed to tunnel into interlayer excitons on ultrafast timescales. Such interlayer excitons have sufficiently long lifetimes that inducing dissociation with external in-plane electric fields becomes an attractive option of improving efficiency of photocurrent devices. In the present paper, we calculate interlayer exciton binding energies, Stark shifts, and dissociation rates for six different transition metal dichalcogenide (TMD) vdWHs using a numerical procedure based on exterior complex scaling (ECS). We utilize an analytical bilayer Keldysh potential describing the interaction between the electron-hole pair, and validate its accuracy by comparing to the full multilayer Poisson equation. Based on this model, we obtain an analytical weak-field expression for the exciton dissociation rate. The heterostructures analysed are MoS$_2$/MoSe$_2$, MoS$_2$/WS$_2$, MoS$_2$/WSe$_2$, MoSe$_2$/WSe$_2$, WS$_2$/MoSe$_2$, and WS$_2$/WSe$_2$ in various dielectric environments. For weak electric fields, we find that WS$_2$/WSe$_2$ supports the fastest dissociation rates among the six structures. We, furthermore, observe that exciton dissociation rates in vdWHs are significantly larger than in their monolayer counterparts. 
\end{abstract}
\begin{document}

\flushbottom
\maketitle
%
%
\thispagestyle{empty}

Naturally occurring layered materials held together by van-der-Waals-like forces have been intensely studied in recent years. Peeling graphite layer-by-layer and forming graphene \cite{Novoselov2004}, a task thought impossible, turned researchers on to two-dimensional materials. Realizing that these layers may be used as building blocks of artificially layered materials, so-called van der Waals heterostructures, provides an endless array of possible combinations \cite{Wang2012,Geim2013}. The extraordinary electronic and optical properties of TMDs \cite{Mak2010,Splendiani2010,Ramasubramaniam2012,Tongay2013,Qiu2013,Gutierrez2013,Mouri2013,Scheuschner2014} have made them one of the most interesting classes of building blocks for vdWHs. Potential electro-optical applications of TMDs include photodetectors \cite{Wang2015,Lopez2013,Yin2012}, light-emitting diodes \cite{Ross2014, Pospischil2014,Withers2015}, and solar cells \cite{Lopez2014,Bernardi2013,Pospischil2014}. It is well known that the optical properties of TMD monolayers are dominated by excitons \cite{Wang2012,Geim2013,Ramasubramaniam2012,Qiu2013,Trolle2015}, as such two-dimensional excitons can have giant binding energies \cite{Ramasubramaniam2012,Olsen2016,Qiu2013,Chernikov2014,Hanbicki2015}. Strongly bound excitons, in turn, make generation of photocurrents difficult, as excitons must first be dissociated into free electrons and holes. This is one of the challenges facing the use of monolayer TMDs in efficient photocurrent devices. A further complication is the fast recombination rates of excitons in these monolayers \cite{Palummo2015,Poellmann2015,Wang2016}. Without inducing dissociation in some way, excitons will typically recombine before they are dissociated. Applying an in-plane electric field to the excitons, however, enhances generation of photocurrents for two reasons: (i) the electric field counteracts recombination by pulling electrons and holes in opposite directions, and (ii) the electric field assists dissociation of excitons \cite{Pedersen2016ExcitonIonization,Haastrup2016,Massicotte2018,Kamban2019}.

When two TMD monolayers are brought together with a type-II band alignment, the conduction band minimum and the valence band maximum reside in two different layers. Electrons and holes in the structure will therefore prefer to reside in separate layers, provided that the loss in exciton binding energy is smaller than the energy gained by band offsets. These spatially indirect electron-hole pairs can still form bound states with large binding energies and they are referred to as interlayer excitons \cite{Hong2014,Chen2016,Miller2017,Kunstmann2018,Merkl2019}. Experiments have shown that photoexcited intralayer excitons created in one of the sheets undergo ultrafast tunneling into interlayer excitons \cite{Merkl2019,Hong2014}. For instance, in MoS$_2$/WS$_2$ heterostructures, it was found that the hole transferred from the MoS$_2$ layer to the WS$_2$ layer within $50$ fs after optical excitation \cite{Hong2014}, and similar time scales were reported for WSe$_2$/WS$_2$ \cite{Merkl2019}. After tunneling, the interlayer excitons have a long lifetime due to the small overlap of electron and hole wave functions \cite{Fogler2014,Rivera2015,Rivera2016,Miller2017,Nagler2017,Jin2018}. Thus, when a weak in-plane electric field is present, the photoexcited intralayer excitons will tunnel into interlayer excitons with sufficiently long lifetimes for them to be dissociated by the electric field. We therefore expect to see significantly larger photocurrents from this type of structure compared to TMD monolayers. This is corroborated by the much larger dissociation rates found for interlayer excitons. As an example, we find that interlayer excitons in freely suspended MoS$_2$/WS$_2$ have a dissociation rate of $\Gamma\approx 1.7\times 10^{4}\, \mathrm{s}^{-1}$ for a field strength of $10\,\mathrm{V/\mu m}$, whereas for excitons in monolayer MoS$_2$ it is only $\Gamma\approx 5.3\times 10^{-38}\, \mathrm{s}^{-1}$. 

The paper is organized as follows. In \cref{sec:II}, we set up the Wannier equation and analyse the interaction between electron-hole pairs in bilayer vdWHs. This is done by solving the multilayer Poisson equation, and subsequently showing that the full solution is excellently approximated by an analytical bilayer Keldysh potential. In \cref{sec:III}, we then turn to computing binding energies as well as Stark shifts and dissociation rates for interlayer excitons. Here, we also compare the numerically exact results to an analytical weak-field approximation, derived by weak-field asymptotic theory. Finally, \cref{sec:IV} concludes upon the results. The solution to the multilayer Poisson equation is presented in \cref{app:poisson}. 

\section{Interlayer excitons}\label{sec:II}
\begin{figure}[t]
	\centering
	\includegraphics[width=1\columnwidth]{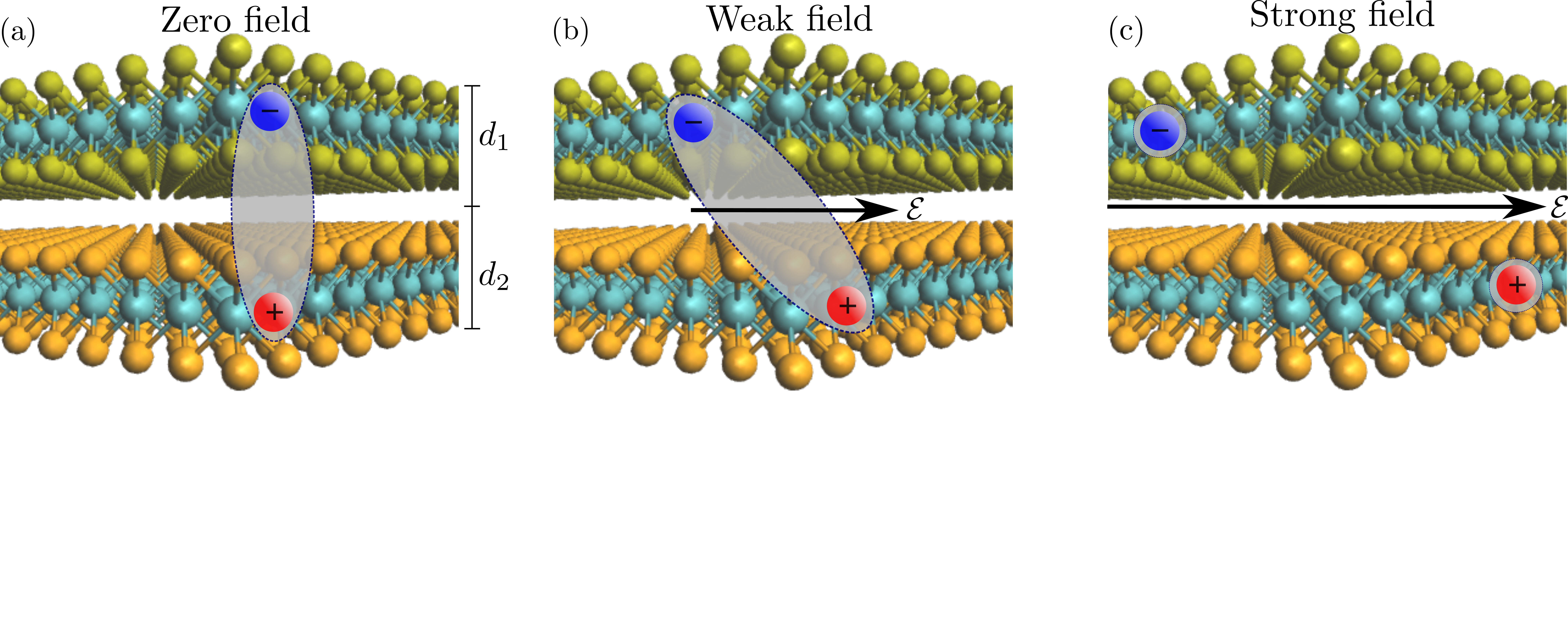}
	\caption{Interlayer exciton in a bilayer van der Waals heterostructure with zero external field (a), a weak in-plane field polarizing the exciton (b), and a strong in-plane field dissociating the exciton (c). }\label{fig:TMDs}
\end{figure}
\begin{figure}[t]
	\centering
	\includegraphics[width=0.5\columnwidth]{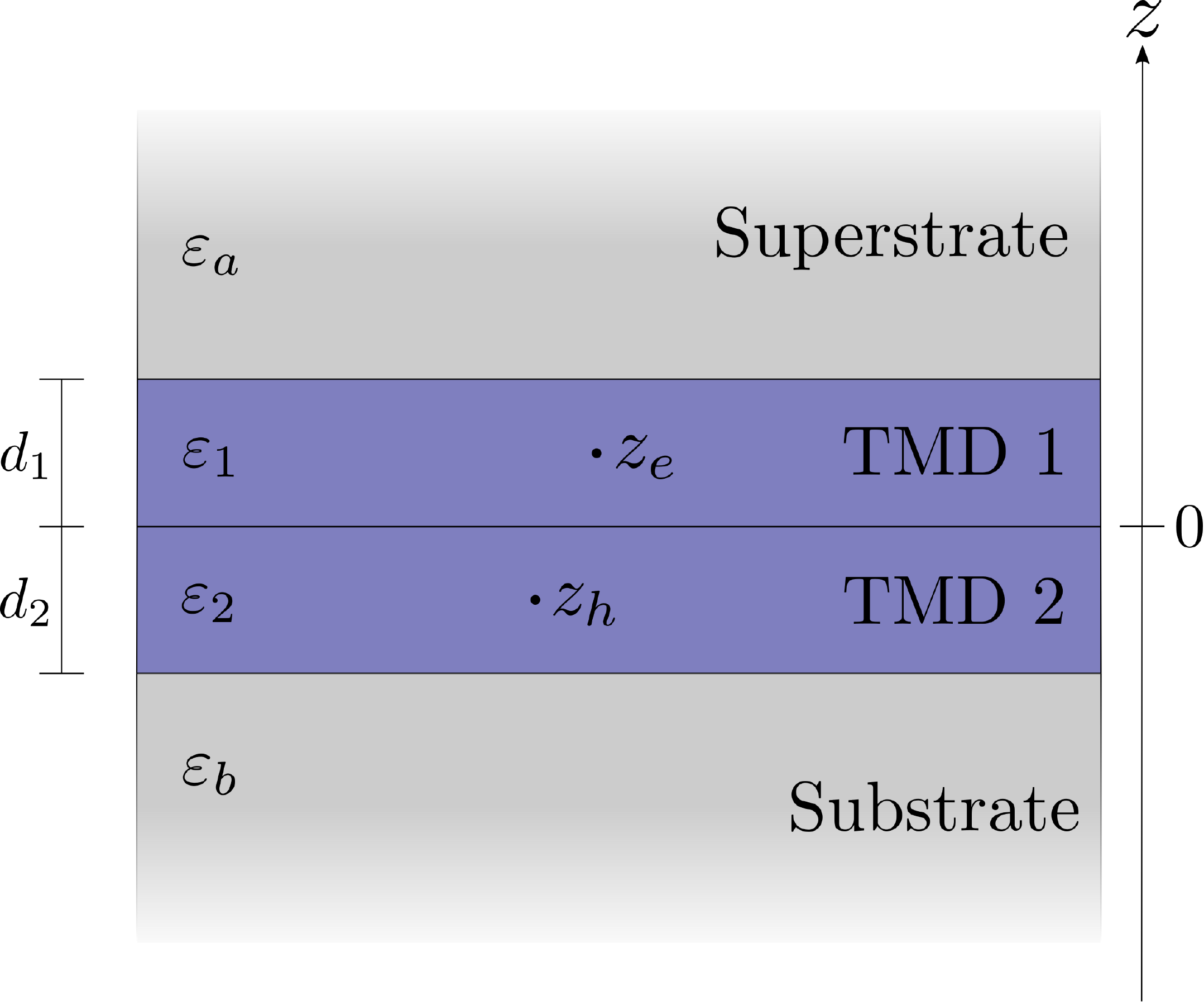}
	\caption{Sketch of the layered geometry used to describe the interaction between an electron at position $z_e$ and a hole at $z_h$. }\label{fig:sketch}
\end{figure}
A bilayer vdWH supports two distinct types of excitons. The electron and hole may either be localized within the same layer, or they may reside in different layers. These two cases are referred to as intra- and interlayer excitons, respectively. The vdWHs considered in the present paper are bilayers with type-II band alignment. We name the structures TMD1/TMD2 so that the conduction band minimum and valence band maximum reside in the first and second layer, respectively. For the structures considered, the energy won from band offsets by the electron and hole residing in the first and second layer, respectively, is larger than the loss in exciton binding energy. The many-body excitonic ground state is therefore an interlayer exciton. Intralayer excitons in either layer are thus excited states, and will therefore tunnel into the ground state. Direct photoexcitation of interlayer excitons by resonance photons is not as likely as for intralayer excitons due to the weak overlap of the electron and hole wave functions. However, resonantly excited intralayer excitons in either layer quickly transition into interlayer excitons \cite{Merkl2019} that, therefore, become very important for many properties of vdWHs. To describe excitonic effects from first principles, one must turn to the many-body Bethe-Salpeter equation \cite{Salpeter1951,Onida2002}. Solving it is  a computationally demanding task even for simple structures. Fortunately, under well-defined approximations, the many-body problem can be simplified to the Wannier equation \cite{Wannier1937,Lederman1976}, essentially reducing it to a Schr\"odinger-type problem with a hydrogenic Hamiltonian. The Wannier model has indeed been shown to yield a sufficiently accurate description of many excitonic properties \cite{Latini2015,Cudazzo2010,Cudazzo2011,Pedersen2016ExcitonStark,Massicotte2018}. In terms of the relative coordinate $\boldsymbol{r} = \boldsymbol{r}_e - \boldsymbol{r}_h$ of the electron-hole pair, it reads (atomic units are used throughout)
\begin{align}
\left[-\frac{1}{2\mu}\nabla^2 +V\left(\boldmath{r}\right) + \pmb{\mathcal{E}}\cdot \boldsymbol{r}\right]\psi\left(\boldsymbol{r}\right)=E\psi\left(\boldsymbol{r}\right)\thinspace,\label{eq:Wannier}
\end{align}
where $\mu = m_em_h/\left(m_e+m_h\right)$ is the reduced exciton mass, $V$ is the screened Coulomb interaction between the electron and hole, and $\pmb{\mathcal{E}}$ is the electric field. The electric field is taken to point along the $x$-axis throughout the paper, i.e. $\pmb{\mathcal{E}} = \mathcal{E}\boldsymbol{e}_x$. As the valence band maximum and conduction band minimum at the $\mathrm{K}$ point are primarily composed of the $d$ orbitals of the metal atoms \cite{Wang2017,Kormanyos2015}, electrons and holes will, to a good approximation, reside in the middle of their respective layers. We are therefore able to freeze their out-of-plane motion, which effectively makes solving \cref{eq:Wannier} a two dimensional problem. \Cref{fig:TMDs} shows an illustration of an interlayer exciton in a bilayer vdWH subjected to three different field strengths: zero (a), weak (b), and strong (c). When an electric field is present, the electron and hole will be pulled in opposite directions. For weak electric fields, the probability of dissociating the exciton is low. It will therefore become polarized, but most likely recombine rather than dissociate. In strong electric fields, however, field induced dissociation becomes likely, and dissociation rates may become extremely large as the field strength increases.

To describe the interaction $V$ between the electron-hole pair, we fix the electron and hole to the middle of their respective sheets. The vertical separation between the electron and hole $d= \left|z_e-z_h\right|$ is therefore $d_{\mathrm{intra}}=0$ and $d_{\mathrm{inter}}=\left(d_1+d_2\right)/2$ for intra- and interlayer excitons, respectively, where $d_1$ and $d_2$ are the thicknesses of the two TMD sheets (see \cref{fig:TMDs}). The van der Waals heterostructure is then modeled as the four layer system in \cref{fig:sketch}. Here, the dielectric function $\varepsilon\left(z\right)$ is taken to be piecewise constant given by $\varepsilon_a$, $\varepsilon_1$, $\varepsilon_2$, and $\varepsilon_b$ in the superstrate, upper TMD sheet, lower TMD sheet, and substrate, respectively. By Fourier decomposing $V$ and solving the multilayer Poisson equation for the Fourier components (see \cref{app:poisson}), one may obtain $V$ expressed as an integral in momentum space
\begin{align}
V\left(r\right)=-\int_0^\infty \frac{e^{-dq}J_0\left(qr\right)}{\varepsilon_{\mathrm{eff}}\left(q\right)}dq\thinspace,\label{eq:fullpotinarticle}
\end{align}
where $J_0$ is a Bessel function. The effective dielectric function $\varepsilon_{\mathrm{eff}}$ is, of course, different for intra- and interlayer excitons. For $q\to 0$, both cases tend to the average dielectric constant of the surrounding media $\left(\varepsilon_a+\varepsilon_b\right)/2$, as expected. For $q\to\infty$, however, the dielectric function describing intralayer excitons tends to the dielectric constant of the layer to which they are confined, i.e. to $\varepsilon_1$ and $\varepsilon_2$ for intralayer excitons in the first and second layer, respectively. On the other hand, the function for interlayer excitons tends to the average dielectric constant of the two layers $\left(\varepsilon_1+\varepsilon_2\right)/2$. The complete interlayer dielectric function is obtained in \cref{app:poisson} and is given by \cref{eq:dielectric}. \Cref{app:poisson} also explains how to obtain the intralayer function. The full potential may readily be obtained for real $r$ by using standard numerical integration techniques in \cref{eq:fullpotinarticle}. However, in the present paper, we seek to calculate dissociation rates by using exterior complex scaling (ECS) \cite{Simon1979,McCurdy1991,Rescigno2000,McCurdy2004}. This implies rotating the radial coordinate into the complex plane outside a radius $R$ by an angle $\phi$, i.e.
\begin{align}
r \to \begin{cases}
r \quad &\mathrm{for }\,\, r<R\\
R + \left(r-R\right)e^{i\phi} \quad &\mathrm{for }\,\, r>R\thinspace.
\end{cases}\label{eq:ECStrans}
\end{align}
It is a simple task to show that the integrand in \cref{eq:fullpotinarticle} will become an exponentially increasing oscillating function when $r> d/\sin\phi+R$ by using the integral representation for the Bessel function \cite{Abramowitz1972}. This makes \cref{eq:fullpotinarticle} extremely difficult to handle while using ECS. Fortunately, the numerical solution is very accurately approximated by a bilayer Keldysh (BLK) potential. The Keldysh potential has been used extensively to describe excitons in monolayer TMDs. The monolayer Keldysh (MLK) interaction is given by \cite{Keldysh1979,Trolle2017}
\begin{align}
V_{\mathrm{MLK}}\left(r\right) = -\frac{\pi}{2r_0}\left[H_0\left(\frac{\kappa r}{r_0}\right)-Y_0\left(\frac{\kappa r}{r_0}\right)\right]\thinspace,\label{eq:monokeld}
\end{align}
where $H_0$ is the zeroth order Struve function \cite{Abramowitz1972}, $Y_0$ is the zeroth order Bessel function of the second kind \cite{Abramowitz1972}, $\kappa = \left(\varepsilon_a+\varepsilon_b\right)/2$ is the average dielectric constant of the surrounding media, and $r_0$ is the screening length proportional to the polarizability of the sheet \cite{Cudazzo2011}. This potential diverges logarithmically at the origin. On the other hand, the potential describing interlayer excitons is finite at the origin due to the vertical separation between the electron-hole pair. 
\begin{figure}[t]
	\centering
	\includegraphics[width=1\columnwidth]{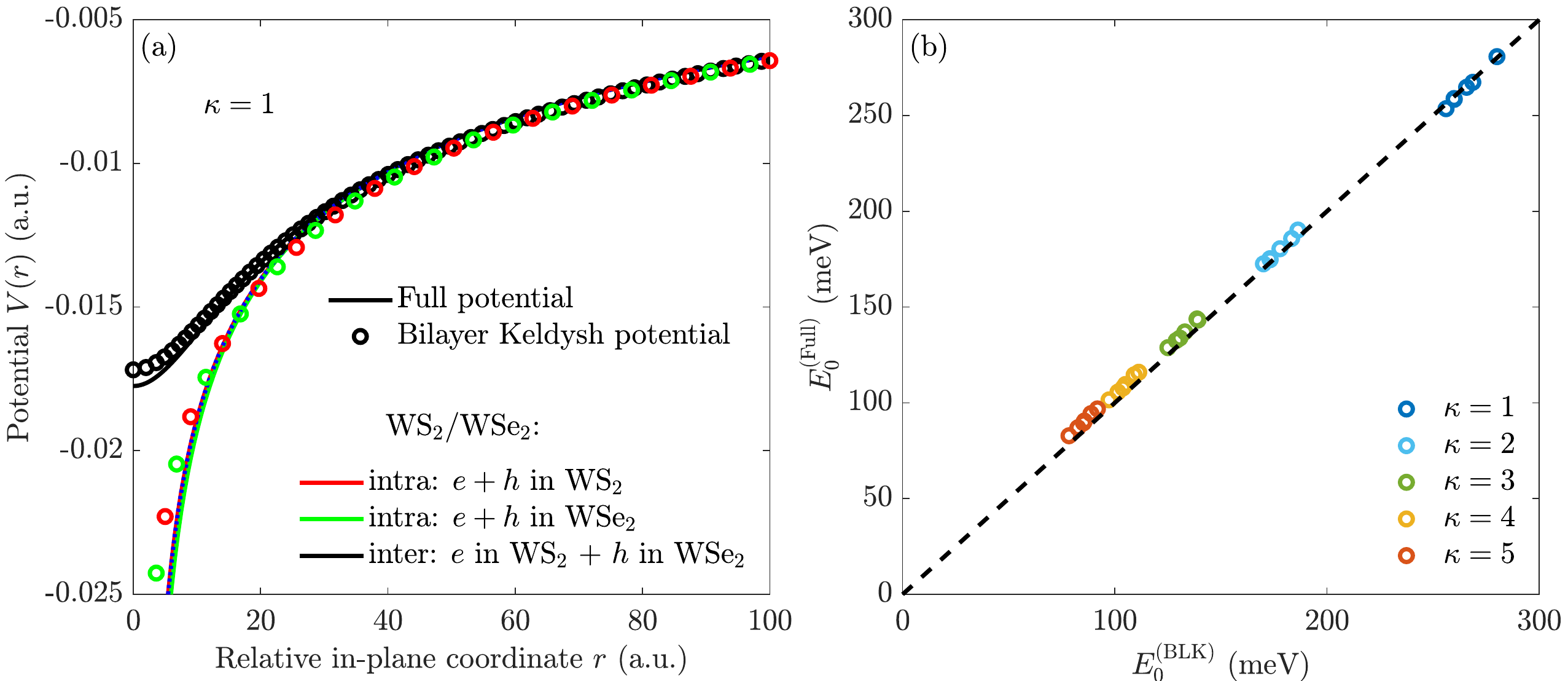}
	\caption{ (a) Exciton Coulomb potential for freely suspended WS$_2$/WSe$_2$. The full  and approximate potentials are shown as the solid lines and circles, respectively. The black line and symbols show the interlayer exciton potential while the red and green lines correspond to the potential of intralayer excitons in the WS$_2$ and WSe$_2$ halves of the bilayer, respectively. (b) Binding energies for interlayer excitons in the six structures obtained with the full potential vs. those obtained by the BLK potential. }\label{fig:potential}
\end{figure}

To obtain the bilayer Keldysh potential, we substitute $r \to \sqrt{r^2 + d^2}$ and $r_0 \to r_0^{\left(1\right)}+r_0^{\left(2\right)}$ into \cref{eq:monokeld} which yields
\begin{align}
V_{\mathrm{BLK}}\left(r\right) = -\frac{\pi}{2\left(r_0^{\left(1\right)}+r_0^{\left(2\right)}\right)}\left[H_0\left(\frac{\kappa \sqrt{r^2 + d^2}}{r_0^{\left(1\right)}+r_0^{\left(2\right)}}\right)-Y_0\left(\frac{\kappa \sqrt{r^2 + d^2}}{r_0^{\left(1\right)}+r_0^{\left(2\right)}}\right)\right]\thinspace.\label{eq:BK}
\end{align}
Here, $r_0^{\left(1\right)}$ and $r_0^{\left(2\right)}$ are the screening lengths of the first and second monolayer, respectively. The first substitution accounts for the possible vertical separation between electrons and holes in bilayer structures. The second substitution accounts for the increased thickness of the structure, as the total screening length is proportional to it \cite{Keldysh1979,Trolle2017}. It should be noted that making the first substitution without the second leads to a potential that is far too strongly binding \cite{Donck2018}. Note, further, that to obtain the interaction for intralayer excitons in a bilayer structure, we use the BLK potential with $d=0$. The screening lengths used in the present paper are \textit{ab initio} values from Ref. \cite{Olsen2016}.
\begin{table}[t]
	\caption{Interlayer exciton binding energy $\left|E_0\right|$ in meV and the field-independent material front factor $\Gamma_0$ of the dissociation rate in atomic units for the six vdWHs in various dielectric environments $\kappa=\left(\varepsilon_a+\varepsilon_b\right)/2$. The reduced interlayer exciton mass $\mu$ is indicated for each heterostructure. }
	\centering
	\begin{adjustbox}{width=1\textwidth}
	\begin{tabular}{c cc c cc c cc c cc c cc c cc}
		\hline\hline\\[-0.2cm]
		\phantom &\multicolumn{2}{ c }{MoS$_2$/MoSe$_2$} 
		&\phantom&\multicolumn{2}{ c }{MoS$_2$/WS$_2$}
		&\phantom&\multicolumn{2}{ c }{MoS$_2$/WSe$_2$}
		&\phantom&\multicolumn{2}{ c }{MoSe$_2$/WSe$_2$}&\phantom&\multicolumn{2}{ c }{WS$_2$/MoSe$_2$}&\phantom&\multicolumn{2}{ c }{WS$_2$/WSe$_2$}\\
		\cmidrule{2-3}  \cmidrule{5-6} \cmidrule{8-9} \cmidrule{11-12} \cmidrule{14-15} \cmidrule{17-18}\\[-0.3cm]
		\phantom & \multicolumn{2}{ c }{$\mu = 0.2636$ }& \phantom & \multicolumn{2}{ c }{$\mu = 0.2039$} &  \phantom& \multicolumn{2}{ c }{$\mu = 0.2039$ }& \phantom & \multicolumn{2}{ c }{$\mu = 0.2221$ }& \phantom& \multicolumn{2}{ c }{$\mu = 0.1862$} &  \phantom& \multicolumn{2}{ c }{$\mu = 0.1543$}\\[0.1cm]
		$\kappa$ &$\left|E_0\right|$  & $\Gamma_0$ & \phantom
		&$\left|E_0\right|$   &$\Gamma_0$ & \phantom
		&$\left|E_0\right|$   &$\Gamma_0$ & \phantom
		&$\left|E_0\right|$ &$\Gamma_0$ & \phantom
		&$\left|E_0\right|$ &$\Gamma_0$ & \phantom
		&$\left|E_0\right|$  &$\Gamma_0$
		\\ [0.1cm]\hline\\[-0.2cm]
		$1$ & $269$ & $1.328\times 10^{-23}$ & \phantom& $280$ & $3.781\times 10^{-20}$ & \phantom& $266$ & $5.616\times 10^{-21}$ & \phantom& $256$ & $1.653\times 10^{-22}$ & \phantom& $260$ & $2.056\times 10^{-20}$ & \phantom& $260$ & $1.265\times 10^{-18}$ \\[0.1cm]
		$2$ & $183$ & $1.089\times 10^{-14}$ & \phantom& $186$ & $7.200\times 10^{-13}$ & \phantom& $178$ & $2.570\times 10^{-13}$ & \phantom& $173$ & $3.925\times 10^{-14}$ & \phantom& $173$ & $5.068\times 10^{-13}$ & \phantom& $170$ & $4.371\times 10^{-12}$ \\[0.1cm]
		$3$ & $139$ & $1.027\times 10^{-11}$ & \phantom& $139$ & $1.864\times 10^{-10}$ & \phantom& $133$ & $9.040\times 10^{-11}$ & \phantom& $131$ & $2.445\times 10^{-11}$ & \phantom& $129$ & $1.444\times 10^{-10}$ & \phantom& $125$ & $6.316\times 10^{-10}$ \\[0.1cm]
		$4$ & $111$ & $3.037\times 10^{-10}$ & \phantom& $109$ & $2.830\times 10^{-9}$ & \phantom& $105$ & $1.609\times 10^{-9}$ & \phantom& $104$ & $5.849\times 10^{-10}$ & \phantom& $101$ & $2.305\times 10^{-9}$ & \phantom& $97$ & $7.091\times 10^{-9}$ \\[0.1cm]
		$5$ & $92$ & $2.226\times 10^{-9}$ & \phantom& $89$ & $1.372\times 10^{-8}$ & \phantom& $86$ & $8.611\times 10^{-9}$ & \phantom& $85$ & $3.759\times 10^{-9}$ & \phantom& $83$ & $1.153\times 10^{-8}$ & \phantom& $78$ & $2.847\times 10^{-8}$ \\[0.1cm]
		
	\end{tabular}\label{tbl:excitonbinding}
\end{adjustbox}
\end{table}
In \cref{fig:potential}, we compare the bilayer Keldysh potential to the full potential obtained by solving the multilayer Poisson equation. Panel (a) shows the potential for the representative WS$_2$/WSe$_2$ case. Evidently, a good agreement is found. Both the intra- and interlayer potentials (exact and approximate) behave as $-1/\kappa r$ for $r\gg d$, as the vertical separation becomes negligible in this region. For small $r$, on the other hand, the intralayer potential diverges logarithmically, while the interlayer potential behaves as $a+br^2$, where $a$ and $b$ are constants. The quadratic form is readily understood by expanding the Bessel function in \cref{eq:fullpotinarticle} to second order, i.e.
\begin{align}
J_0\left(qr\right) = 1-\left(\frac{qr}{2}\right)^2 + O\left(r^4\right)\thinspace.
\end{align}
Integrating using the second order expansion shows that the integral diverges for $d=0$. For $d>0$, on the other hand, the exponential function leads to converging integrals even for $r=0$. The quadratic behaviour has inspired the use of a harmonic oscillator model, obtained by expanding a potential similar to \cref{eq:BK} to second order, to analyse interlayer excitons \cite{Berman2017}. It should be noted, however,  that the binding energies predicted by a second order expansion of \cref{eq:BK} are in poor agreement with those obtained using the full potential.  In contrast, the binding energies obtained with the BLK potential are in good agreement with the full potential results, as seen in panel (b) of \cref{fig:potential}. In fact, they never deviate more than $6\%$ for the cases considered. The binding energies are, furthermore, in excellent agreement with those found in literature. As an example, the binding energy of $1$s interlayer excitons in WS$_2$/WSe$_2$ on a diamond substrate ($\varepsilon_a = 1$ and $\varepsilon_b = 5$) was measured recently to be $126\pm7$ meV \cite{Merkl2019}, and our model yields $125$ meV. Furthermore, Ovesen \textit{et al.} \cite{Ovesen2019} found the binding energy of interlayer excitons in MoSe$_2$/WSe$_2$ in free space to be $246$ meV using a model similar to our full potential, where we find $260$ meV with the BLK model. It should be mentioned that the model used in the present paper predicts slightly lower free-space binding energies than some of the \textit{ab initio} methods \cite{Meckbach2018,Gillen2018,Torun2018,Deilmann2018}.  Interlayer exciton binding energies for the six TMD bilayer combinations with type-II band alignment \cite{Gong2013,Komsa2013} are summarized in \cref{tbl:excitonbinding} for various dielectric environments. The effective masses used are obtained from Ref. \cite{Kormanyos2015}, and the TMD widths have been taken as half the vertical lattice constants found in Ref. \cite{He2014}.

\section{Field induced dissociation}\label{sec:III}
\begin{figure}[t]
	\centering
	\includegraphics[width=1\columnwidth]{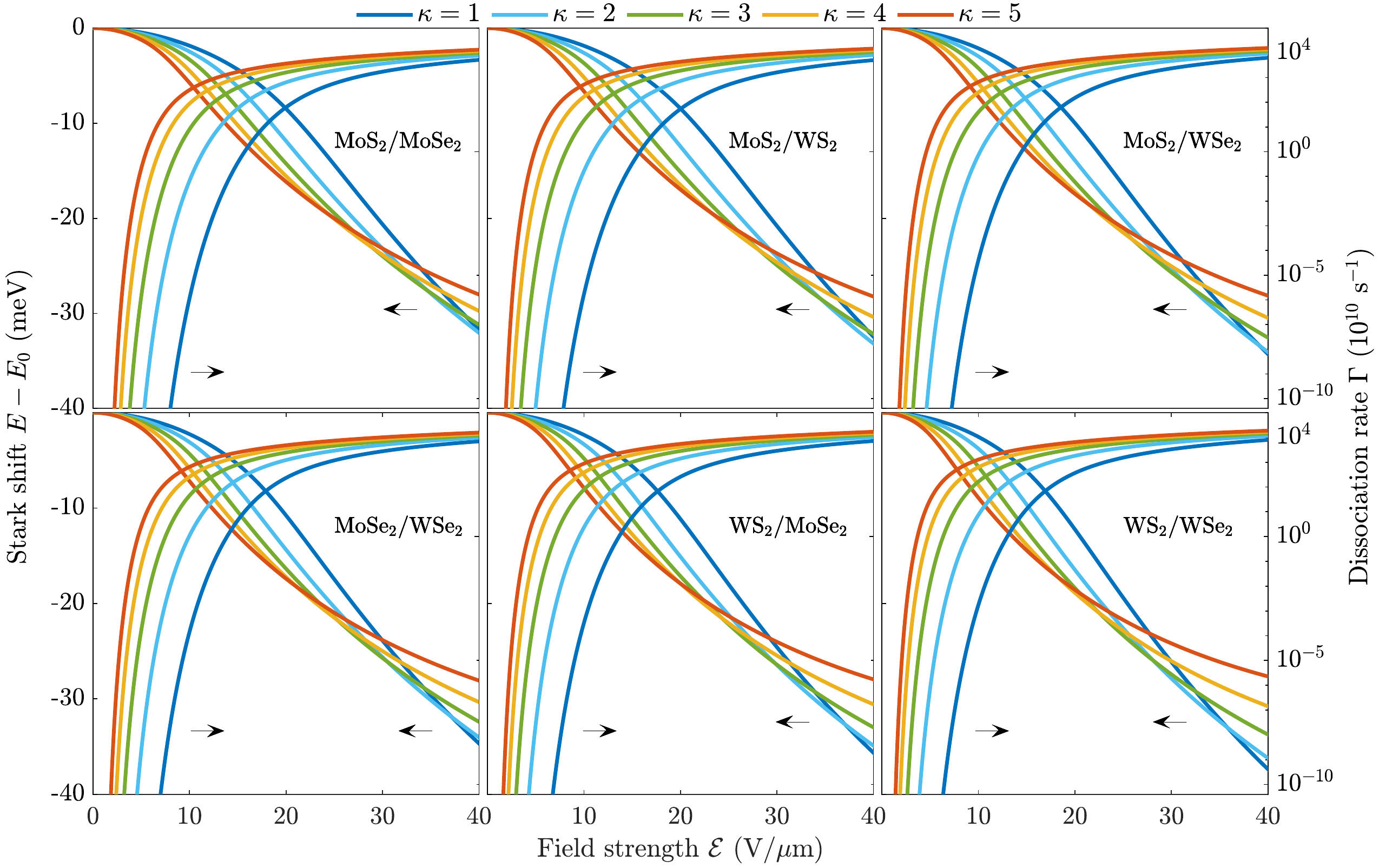}
	\caption{Stark shift (left axis) and dissociation rate (right axis) of interlayer excitons in six different bilayer van der Waals heterostructures in various dielectric environments $\kappa$. }\label{fig:dissociation}
\end{figure}




When the exciton is subjected to an electrostatic field, it may be dissociated. This is realized in the Wannier model by the energy eigenvalue obtaining a non-vanishing imaginary part \cite{Pedersen2016ExcitonIonization,Pedersen2016Stark,Haastrup2016,Massicotte2018}. The field induced dissociation rate is then given by $\Gamma = -2 \,\mathrm{Im}\, E$, where the imaginary part can be obtained efficiently by utilizing the ECS technique \cite{Simon1979,McCurdy1991,Rescigno2000,McCurdy2004}. As mentioned briefly in the previous section, this technique consists of rotating the radial coordinate into the complex plane outside a radius $R$ (see \cref{eq:ECStrans}). The partitioning of the radial domain is efficiently dealt with by resolving the radial part of the eigenstate in a finite element basis consisting of Legendre polynomials, and the angular part in a cosine basis. We have previously used the same numerical procedure to calculate dissociation rates for excitons in various monolayer TMDs \cite{Kamban2019}, and we refer the interested reader to that paper for the technical details of the method. The field induced dissociation rates and Stark shifts for interlayer excitons in the six van der Waals heterostructures are shown in \cref{fig:dissociation}. As is evident, the structures support excitons that behave very similarly in electrostatic fields. The exciton Stark shifts can be seen to vary approximately as $\mathcal{E}^2$ for weak electric fields, in accordance with perturbation theory $E \approx E_0 - \frac{1}{2}\alpha\mathcal{E}^2$, where $E_0$ is the unperturbed energy and $\alpha$ the in-plane polarizability. Calculating the polarizability using the Dalgarno-Lewis equation \cite{Dalgarno1955} and a finite element basis \cite{Kamban2019} reveals that the interlayer exciton polarizabilities are significantly larger than their monolayer counterparts. For example, freely suspended MoS$_2$/WSe$_2$ supports interlayer excitons with $\alpha_{\mathrm{MoS}_2/\mathrm{WSe}_2}^{\left(\kappa=1\right)} = 41\times 10^{-18} \mathrm{eV}\left(\mathrm{m/V}\right)^2$, whereas monolayer MoS$_2$ and WSe$_2$ support exciton polarizabilities of around $\alpha_{\mathrm{MoS}_2}^{\left(\kappa=1\right)} = 4.6\times 10^{-18} \,\mathrm{eV}\left(\mathrm{m/V}\right)^2$ and $\alpha_{\mathrm{WSe}_2}^{\left(\kappa=1\right)} = 6.3\times 10^{-18}\, \mathrm{eV}\left(\mathrm{m/V}\right)^2$, respectively \cite{Pedersen2016ExcitonStark}. Encapsulating the materials in hBN ($\kappa = 4.9$ \cite{Latini2015}) increases the polarizabilities to $\alpha_{\mathrm{MoS}_2/\mathrm{WSe}_2}^{\left(\kappa=4.9\right)} = 116\times 10^{-18}\, \mathrm{eV}\left(\mathrm{m/V}\right)^2$ whereas $\alpha_{\mathrm{MoS}_2}^{\left(\kappa=4.9\right)} = 14.2\times 10^{-18}\, \mathrm{eV}\left(\mathrm{m/V}\right)^2$ and $\alpha_{\mathrm{WSe}_2}^{\left(\kappa=4.9\right)} = 20.8\times 10^{-18}\, \mathrm{eV}\left(\mathrm{m/V}\right)^2$  \cite{Pedersen2016ExcitonStark}. The reason that these large polarizabilities are observed for interlayer excitons in bilayer vdWHs is the increased screening and the vertical separation of the electron and hole. Both reduce the binding energy and they are therefore much easier to polarize.  
\begin{figure}[t]
	\centering
	\includegraphics[width=0.5\columnwidth]{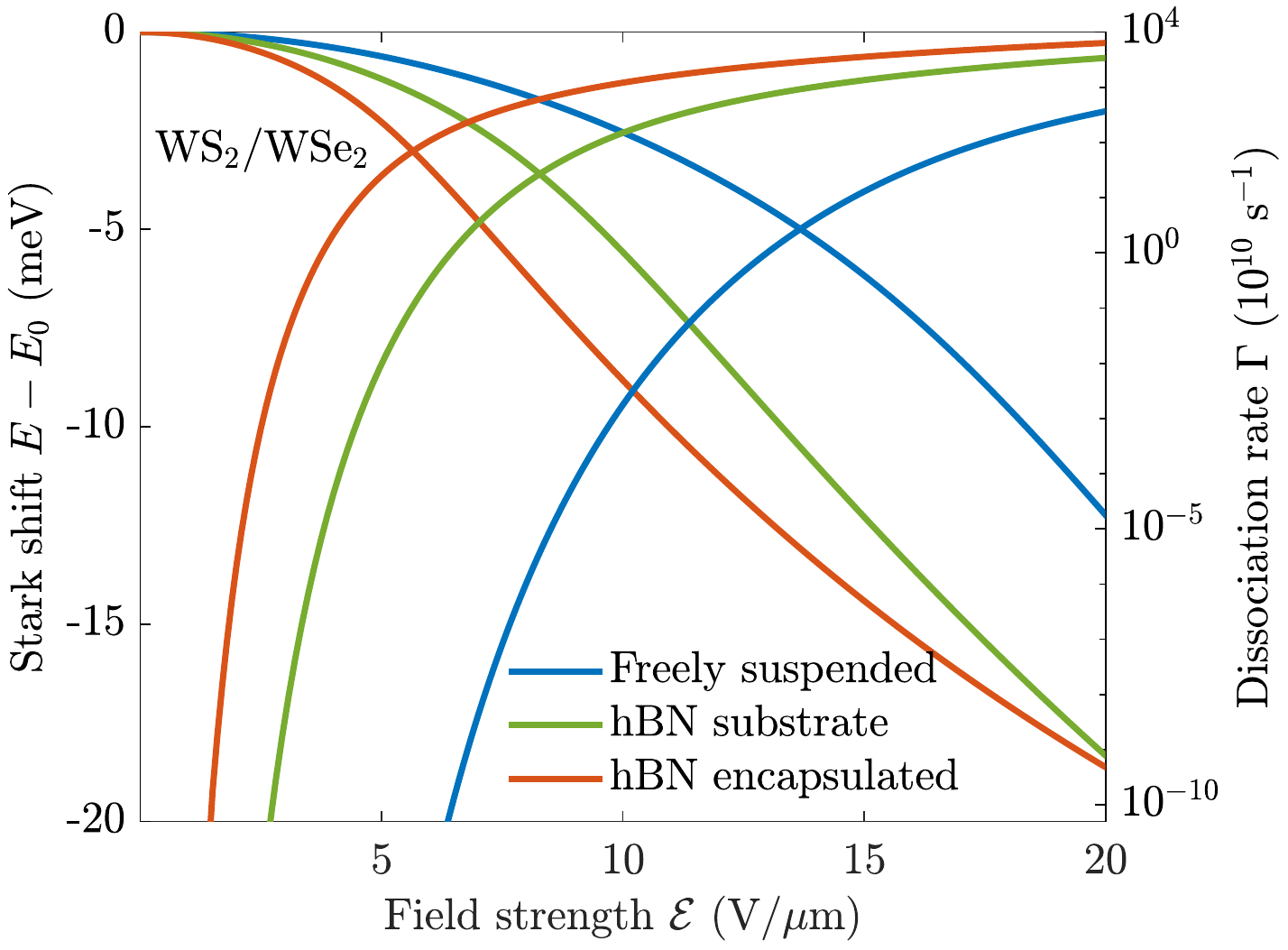}
	\caption{Stark shift (left axis) and dissociation rate (right axis) for interlayer excitons in a WS$_2$/WSe$_2$ bilayer that is either (i) freely suspended, (ii) placed on an hBN substrate, or (iii) encapsulated by hBN. }\label{fig:dissWSe2WS2}
\end{figure}

Turning to the dissociation rates, it is clear that the encapsulating media significantly alter how quickly interlayer excitons are dissociated. The same behaviour was observed for their monolayer counterparts \cite{Kamban2019}, and one therefore has a large degree of freedom in device design. For extremely weak fields, the dissociation rates are very low, but they grow rapidly with an increasing field strength. As an example, the dissociation rate of interlayer excitons in freely suspended MoS$_2$/WS$_2$ is around $\Gamma\approx 1.7\times 10^{4}\, \mathrm{s}^{-1}$ already at $\mathcal{E} = 10\,\mathrm{V/\mu m}$, and only $\Gamma\approx 5.3\times 10^{-38}\, \mathrm{s}^{-1}$ and $\Gamma\approx 2.7\times 10^{-33}\, \mathrm{s}^{-1}$ for monolayer MoS$_2$ and WS$_2$, respectively \cite{Kamban2019}. It should be noted that dissociation rates of intralayer excitons are only important if they are comparable to the rate at which intralayer excitons tunnel over to interlayer excitons. In a recent experiment on MoS$_2$/WS$_2$ structures, the holes of photoexcited excitons in the MoS$_2$ layer of this structure were observed to tunnel into the WS$_2$ layer within $50\, \mathrm{fs}$ \cite{Hong2014}. Comparing to the dissociation rates of intralayer excitons in the top and bottom layer of MoS$_2$/WS$_2$ for $\mathcal{E}=10\,\mathrm{V/\mu m}$, we find $\Gamma\approx 2.1\times 10^{-3}\, \mathrm{s}^{-1}$ and $\Gamma\approx 2.9\times 10^{4}\, \mathrm{s}^{-1}$, respectively. The large difference between these two rates can be traced back to the reduced masses, which are $\mu = 0.2513$ and $\mu = 0.1543$, respectively. The time it takes to dissociate these excitons with the given field may be approximated as $\tau = 1/\Gamma \approx 476\, \mathrm{s}$ and $\tau \approx 3.5\times 10^{-5}\, \mathrm{s}$, respectively. The intralayer excitons have therefore clearly transitioned to interlayer excitons before they are dissociated by the field. Interlayer tunneling rates are likely affected by material parameters as well as the surrounding media. Assuming, however, similar time scales as observed for MoS$_2$/WS$_2$ $\Gamma_{\mathrm{Tunnel}}\approx 10^{13}\, \mathrm{s}^{-1}$, interlayer dissociation rates are the limiting factor in field induced exciton dissociation for weak to moderate fields. For the largest fields in \cref{fig:dissociation}, the competition between tunneling and dissociation will be important for an accurate description. Due to the risk of dielectric break-down, such large fields are best avoided in devices, however. The high interlayer dissociation rates suggest that using carefully chosen bilayer TMDs in photocurrent devices is much more attractive than their monolayer counterparts. Moreover, proper encapsulation will further improve device performance. As hBN is a very common material used to encapsulate samples, we show the interlayer dissociation rates for WS$_2$/WSe$_2$ that is either (i) freely suspended, (ii) placed on an hBN substrate, or (iii) encapsulated by hBN in \cref{fig:dissWSe2WS2}. Evidently, hBN surroundings increase the dissociation rates by several orders of magnitude, and for weak fields in particular.

\begin{figure}[t]
	\centering
	\includegraphics[width=1\columnwidth]{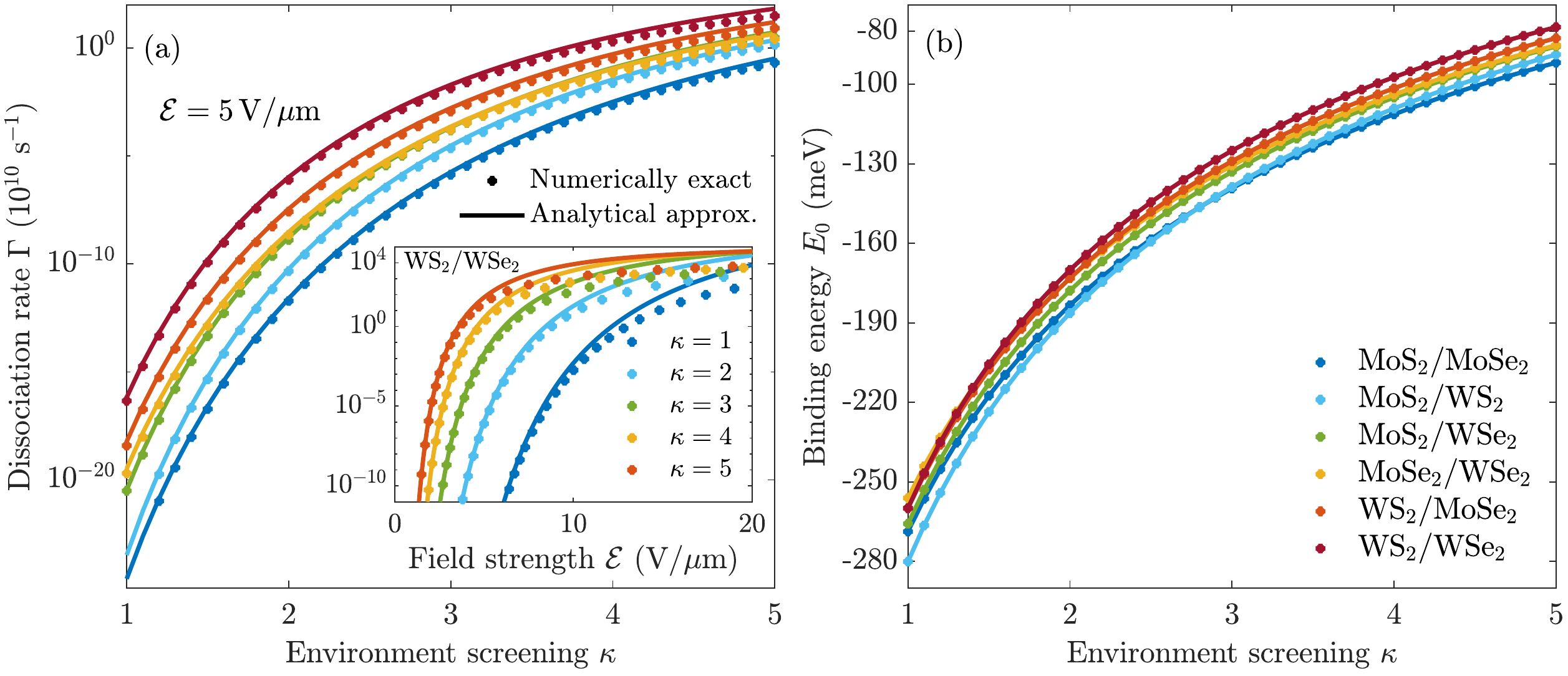}
	\caption{Dissociation rates (a) and binding energies (b) for interlayer excitons in six vdWHs as functions of environment screening $\kappa$. In panel (a), the dots and solid lines represent the numerically exact and analytical approximation \cref{eq:wfexp}, respectively. The inset shows the approximate and exact dissociation rates for interlayer excitons in WS$_2$/WSe$_2$ as a function of field strength. }\label{fig:dissvskappa}
\end{figure}

Recently an analytical weak-field approximation was obtained for exciton dissociation rates in monolayer TMDs \cite{Kamban2019}. The derivation was made using weak-field asymptotic theory \cite{Tolstikhin2011}, and the fact that the MLK potential has a simple asymptotic form. As mentioned in the previous section, the BLK potential has exactly the same asymptotic behaviour. The weak-field approximation for interlayer excitons therefore has exactly the same form, albeit with a different field-independent front factor. We arrive at
\begin{align}
\Gamma \approx \Gamma_0\mathcal{E}^{1/2-2\sqrt{\mu}/\left(\kappa k\right)}\exp\left(-\frac{2\sqrt{\mu}k^3}{3\mathcal{E}}\right)\thinspace,\label{eq:wfexp}
\end{align}
where $k = \sqrt{2\left|E_0\right|}$ and $\Gamma_0 = \Gamma_0\left(E_0,\mu,r_0^{\left(1\right)}+r_0^{\left(2\right)},d,\kappa\right)$ is a field-independent material constant. The parameters needed to use \cref{eq:wfexp} are presented in \cref{tbl:excitonbinding}, where $\Gamma_0$ has been computed by the integral procedure in \cite{Kamban2019}. Panel (a) of \cref{fig:dissvskappa} shows the interlayer dissociation rates as functions of environment screening $\kappa$ for a field of $5\,\mathrm{V/\mu m}$ computed with the numerical procedure (dots) and the weak-field approximation (solid lines). Evidently a good agreement is found for such a low field. Nevertheless, as the inset shows, the weak-field approximation quickly begins to overestimate the dissociation rate for larger fields. This was also found for monolayer TMDs \cite{Kamban2019}. It is clear that WS$_2$/WSe$_2$ supports excitons with the largest dissociation rates at $\mathcal{E}=5\,\mathrm{V/\mu m}$, making it an interesting choice in device design. Note that the fully numerical procedure breaks down if the dissociation rate becomes extremely small, and, hence, we are unable to use it to obtain dissociation rates for MoS$_2$/MoSe$_2$ and MoS$_2$/WS$_2$ in surroundings with (very) low screening for this field strength. It is therefore advantageous to have a formula such as \cref{eq:wfexp} when the fields become sufficiently weak. 

In panel (b) of \cref{fig:dissvskappa}, we show the binding energies for the same excitons as functions of $\kappa$. They clearly follow similar trends as the dissociation rates do, suggesting that binding energy has a significant impact on dissociation rates. This is to be expected, as strongly bound excitons are harder to pull apart. It should, however, be noted that the binding energy does not uniquely determine the dissociation rate. As an example, the interlayer excitons in MoS$_2$/MoSe$_2$ and MoS$_2$/WS$_2$ have identical binding energies for surrounding media with $\kappa\approx 2.8$. Nevertheless, the dissociation rates are clearly different. In fact, several crossings are observed in the binding energies as a function of $\kappa$ whereas only one is found in the dissociation rates at $5\,\mathrm{V/\mu m}$. The origin is the different reduced exciton masses. For no structure does $\Gamma_0$ nor $\sqrt{\mu}/k$ contain any crossing for $\kappa\in\left[1;5\right]$. However, $\sqrt{\mu}k^3$ crosses for MoS$_2$/WSe$_2$ and MoSe$_2$/WSe$_2$. This complicated behaviour suggests that, to obtain accurate estimates of interlayer exciton dissociation rates, one must turn to either the weak-field approximation or make a full numerical calculation. 



\section{Conclusion}\label{sec:IV}

In the present paper, we have studied binding energies, Stark shifts, and dissociation rates of interlayer excitons in van der Waals heterostructures (vdWHs). The structures analysed are the six bilayer vdWHs with type-II band alignment arising from combinations of MoS$_2$, MoSe$_2$, WS$_2$, and WSe$_2$. The bilayer excitons are described using an analytical bilayer Keldysh potential, which we have verified for accuracy by comparing to the full solution of the multilayer Poisson equation. Exciton binding energies, Stark shifts, and dissociation rates can therefore readily be calculated using the analytical potential. We find interlayer exciton binding energies ranging from $256$ to $280\,\mathrm{meV}$ for freely suspended structures and from $78$ to $92\,\mathrm{meV}$ for heavily screened structures, making them stable at room temperature. Furthermore, both the polarizabilities and dissociation rates found for these excitons are much larger than their monolayer counterparts. For example, interlayer excitons in freely suspended MoS$_2$/WS$_2$ are found to dissociate at a rate of $\Gamma\approx 1.7\times 10^{4}\, \mathrm{s}^{-1}$ in a field strength of $10\,\mathrm{V/\mu m}$ whereas monolayer MoS$_2$ and WS$_2$ have $\Gamma\approx 5.3\times 10^{-38}\, \mathrm{s}^{-1}$ and $\Gamma\approx 2.7\times 10^{-33}\, \mathrm{s}^{-1}$, respectively. 

For moderate field strengths, intralayer exciton dissociation rates are significantly lower than the rate at which such excitons tunnel into interlayer excitons. For this reason, interlayer exciton dissociation rates are the limiting factor in generation of photocurrents at weak to moderate fields. Since optically excited excitons in one of the layers tunnel to long-lived interlayer excitons on ultrafast timescales, bilayer vdWHs with favourable band offsets may potentially serve as building blocks in efficient photocurrent devices. Finally, the numerically exact dissociation rates are compared to an analytical weak-field dissociation formula obtained from weak-field asymptotic theory. A good agreement is found in the weak-field limit, and \cref{eq:wfexp} therefore serves as a useful formula to quickly estimate field induced dissociation rates of excitons in bilayer vdWHs for weak electric fields. 
\appendix
\section{Multilayer Poisson equation}\label{app:poisson}

We want to find the interaction between the two particles in the system represented by \cref{fig:sketch}. For charges at $z$ and $z'$ with in-plane separation $\boldsymbol{r}$, the interaction can be Fourier decomposed as
\begin{align}
-V\left(r,z,z'\right) = \frac{1}{4\pi^2}\int \varphi\left(z,z';q\right)e^{i\boldsymbol{q}\cdot \boldsymbol{r}}d^2q\thinspace,
\end{align}
where the Fourier components satisfy the Poisson equation
\begin{align}
4\pi\delta\left(z-z'\right) = \left[q^2\varepsilon\left(z;q\right)-\frac{\partial}{\partial z}\varepsilon\left(z;q\right)\frac{\partial}{\partial z}\right]\varphi\left(z,z';q\right)\thinspace.
\end{align}
We take the dielectric function $\varepsilon$ to be piecewise constant
\begin{align}
\varepsilon\left(z;q\right) = \begin{cases}
\varepsilon_a\thinspace, \quad  z>d_1\\
\varepsilon_1\thinspace, \quad 0<z<d_1\\
\varepsilon_2\thinspace, \quad -d_2<z<0\\
\varepsilon_b\thinspace, \quad  z<-d_2\thinspace,
\end{cases}
\end{align}
where $d_1$ and $d_2$ are the widths of the first and second layer, respectively. For $z'$ confined to the second layer, the solution can be sought on the form
\begin{align}
\varphi\left(z,z';q\right) = \frac{2\pi}{q}\begin{cases}
A_1e^{-qz}\\
A_2e^{-qz} + B_2e^{qz}\\
A_3e^{-qz} + B_3e^{qz} +\varepsilon_2^{-1}e^{-q\left|z-z'\right|}\\
B_4e^{qz}\thinspace,
\end{cases}
\end{align}
in the respective regions. The Fourier components can then be found analytically by solving the system of equations that arises from the boundary conditions. To describe charges confined to distinct layers, we fix the electron and hole to $z=d_1/2$ and $z'=-d_2/2$, respectively. This leads to the interlayer exciton potential
\begin{align}
\varphi\left(d_1/2,-d_2/2;q\right) = \frac{\varphi_0\left(d_1/2,-d_2/2;q\right)}{\varepsilon_{\mathrm{eff}}\left(q\right)}\thinspace,
\end{align}
where
\begin{align}
\varphi_0\left(z,z';q\right) = \frac{2\pi}{q}e^{-\left|z-z'\right|q}
\end{align}
is the bare interaction. The effective dielectric function is given by
\begin{align}
\varepsilon_{\mathrm{eff}}\left(q\right) = \frac{A\left(q\right)}{B\left(q\right)}\label{eq:dielectric}.
\end{align}
with
\begin{align}
A\left(q\right) = \left(1-e^{-2d_1q}\right)\left(\varepsilon_1^2\gamma^+ + \varepsilon_a\varepsilon_2\gamma^-\right) + \left(1+e^{-2d_1q}\right)\left(\varepsilon_a\varepsilon_1\gamma^+ + \varepsilon_1\varepsilon_2\gamma^-\right)\thinspace,
\end{align}
\begin{align}
B\left(q\right) = 2\left[\left(1-e^{-d_1q}\right)\varepsilon_a + \left(1+e^{-d_1q}\right)\varepsilon_1\right] \times \left[\left(1+e^{-d_2q}\right)\varepsilon_2+\left(1-e^{-d_2q}\right)\varepsilon_b\right]\thinspace,
\end{align}
where
\begin{align}
\gamma^\pm = \left(1\pm e^{-2d_2q}\right)\varepsilon_2 + \left(1\mp e^{-2d_2q}\right)\varepsilon_b\thinspace.
\end{align}
This effective dielectric function tends to $\left(\varepsilon_a+\varepsilon_b\right)/2$ for $q\to 0$ and to $\left(\varepsilon_1+\varepsilon_2\right)/2$ for $q\to \infty$, as expected. The dielectric function describing charges confined to the same layer may be obtained in a similar manner by placing both charges at $z=-d_2/2$. The interaction in real space can then be obtained as the inverse Fourier transform
\begin{align}
-V\left(r\right)=\int_0^\infty \frac{e^{-\left|z-z'\right|q}J_0\left(qr\right)}{\varepsilon_{\mathrm{eff}}\left(q\right)}dq\thinspace.\label{eq:fullpot}
\end{align}
For the dielectric constants, we have used the static in-plane dielectric constants calculated from first principles in Ref. \cite{Laturia2018}.

\bibliography{litt}

\section*{Acknowledgements}

The authors gratefully acknowledge financial support from the Center for Nanostructured Graphene (CNG), which is sponsored by the Danish National Research Foundation, Project No. DNRF103. Additionally, T.G.P. is supported by the QUSCOPE Center, sponsored by the Villum Foundation.

\section*{Author contributions statement}

H.C.K. initiated the work, performed all calculations, and prepared the initial manuscript. T.G.P. assisted in solving the Poisson equation. H.C.K. and T.G.P. analysed the results and prepared the final manuscript. 

\section*{Additional information}

\textbf{Competing interests} The authors declare no competing financial interest. 

\end{document}